\documentclass[prc,showpacs,twocolumn]{revtex4}
\usepackage{graphicx}

\begin{document}

\title{Gamow-Teller transitions and deformation in the proton-neutron random phase approximation}

\author{Ionel Stetcu}
\altaffiliation{Present address: Department of Physics, University of Arizona,
PO Box 210081, Tucson, AZ 85721.}
\affiliation{
Department of Physics and Astronomy,
Louisiana State University,
Baton Rouge, Louisiana 70803-4001}
\author{Calvin W.~Johnson}
\affiliation{Physics Department,
San Diego State University,
5500 Campanile Drive, San Diego, California 92182-1233}

\begin{abstract}
We investigate  reliability of Gamow-Teller transition
strengths computed in the proton-neutron random phase
approximation, comparing with exact results from diagonalization
in full $0\hbar\omega$ shell-model spaces. By  allowing the
Hartree-Fock state to be deformed, we obtain good results for a
wide variety of nuclides, even though we do not project onto good
angular momentum.  We suggest that deformation is as important or
more so than pairing for Gamow-Teller transitions.
\end{abstract}
\maketitle

\section{Introduction}

Weak processes, such as beta ($\beta$) and
double-beta ($\beta\beta$) decay, have deep
consequences for nucleosynthesis \cite{LMP03} and physics beyond the Standard
Model \cite{Zde02}. Weak processes are also sensitive to details of nuclear
structure: allowed Gamow-Teller (GT) transitions depend in
particular upon Pauli blocking \cite{LMP03,AZZ93}.  Thus, predictions for
astrophysics as well as interpretation of $\beta\beta$-decay
experiments must be modeled with care. Because of this sensitivity
to Pauli blocking and thus upon configuration mixing, the large-basis
interacting shell model (SM) \cite{BW88} provides one of the best microscopic
approaches to Gamow-Teller transitions.

The interacting shell model, however, is computationally
expensive, and only recently has the full $0 \hbar \omega \, \,
pf$-shell become tractable. A simpler approach is the random phase
approximation (RPA) and its generalizations, which have been
widely and successfully applied to giant resonances such as the
electric dipole (giant dipole resonance or GDR) \cite{Goeke82},
and has also been applied to many important problems in weak
transitions \cite{LMP03,jameel,c12rpa,2betarpa,ZB93,engel1997}. It
is not immediately obvious, however, that the RPA is an
appropriate approximation for \textit{all} transitions. Because
the RPA is the small-amplitude limit of time-dependent mean-field
theory \cite{Rowe70,ring}, it seems appropriate for the GDR, which
can be described semiclassically in the Goldhaber-Teller model
\cite{GT48} as protons oscillating in bulk against neutrons. The
application of the RPA and its extensions to Gamow-Teller
transitions, although with a long history \cite{KS63,pnQRPA},  is
more problematic. Two implicit assumptions in the RPA are, first,
ground-state correlations on top of a mean-field state are small,
and, second, particle-hole phonons have boson commutation rules,
which means that Pauli blocking is not fully treated.  Therefore,
as GT transitions are sensitive to Pauli blocking, they may not be
well matched to RPA calculations.

Reading the literature only furthers these doubts. A number of
authors have previously tested the efficacy of calculating GT
transitions within the proton-neutron quasiparticle RPA (pnQRPA),
through comparison with exact calculations either with full
shell-model diagonalization \cite{Lau88,CMS91,ZB93,ABB93} or
group-theoretical schematic models \cite{engel1997}.  The efficacy
of these pnQRPA calculations, which we will discuss in more detail
below, can be broadly summarized as poor, typically overestimating
the total transition strength and underestimating the first moment
of the transition strength. In order to compare with the
calculations described below, is important to understand that
these calculations used spherical $J=0$, $N$-even $Z$-even parent
states and treated pairing carefully, starting with either the
Hartree-Fock-Bogoliubov or the Bardeen-Cooper-Schrieffer
equations.

With the exception of those tests of Gamow-Teller transitions,
however, most other tests of the RPA and its generalizations have
been against toy models. Recently we began to systematically test
the RPA against full shell-model diagonalization
\cite{stetcu2002,johnson2002,stetcu2003}. In this paper we
describe the generalization to proton-neutron RPA (pnRPA) and
compare Gamow-Teller transitions against the full shell-model
results. In contrast to previous approaches, we do not treat
pairing carefully, but do allow arbitrary deformation, even
though we do not project to good angular momentum. Furthermore we
are not limited to even-even nuclides.
 We obtain good transition
strengths, and where we can compare to published spherical pnQRPA
strengths, our calculations are generally superior. We therefore
conclude that proper treatment of deformation, even without
projection, is at least as important as proper treatment of pairing,
and arguably more so.
This agrees with shell model studies that show a correlation
between E2 transition strengths (a measure of deformation) and GT
transition strengths \cite{AZZ93,TDH96}.

We also briefly argue that the pnRPA frequencies should be real,
and not complex; such an argument is missing from the
literature. Included in this discussion is a lemma
helpful to understanding the solution of the pnRPA eigenvalue
equations.

\section{Formalism}

 In this paper we consider only
change-changing Gamow-Teller transitions; the transition operators
are thus:
\begin{equation}
{\cal O}_\pm = g_A\vec{\sigma} \tau_\pm,
\label{defOp}
\end{equation}
where $\tau_{+}$ changes a neutron into a proton. Because here we consider only strength
\textit{distributions} and not absolute transition strengths, we
drop the axial vector coupling $g_A$.
The transition strength from the parent state $| i
\rangle$ (here always a ground state) to a final state $f$
at  excitation energy
$E = E_f - E_i$ is given by
\begin{equation}
S(E) = \sum_f \delta(E_f -E_i - E) \left | \left \langle f \right |
{\cal O} \left | i \right \rangle \right |^2.
\label{trStr}
\end{equation}
The transition strength to a given level is also often called the
B(GT) value.

A convenient way to characterize the distribution of transition
strength is through moments. The zeroth moment is the total
transition strength, $S_0$,
\begin{equation}
S_0 = \sum_{E} S(E),
\end{equation}
while the \textit{centroid} of the distribution $\bar{E}$ is the first
moment
\begin{equation}
\bar{E} = \sum_{E}E \, S(E)/S_0
\end{equation}
(note that this centroid is relative to the parent energy $E_i$)
and the \textit{width} of the distribution, $\Delta E$,  is the
square root of the second moment,
calculated relative to the centroid:
\begin{equation}
\Delta E^2 = \sum_{E}(E-\bar{E})^2 S(E)/S_0.
\end{equation}

An important check of any calculation is the well-known Ikeda sum
rule, which holds true for any parent state, both in the SM and in the
RPA:
\begin{equation}
S_0(\beta^-) - S_0(\beta^+) = 3(N-Z).
\label{IkedaSR}
\end{equation}

We now briefly review the random phase approximation and its
variant, the proton-neutron RPA (pnRPA) \cite{pnQRPA}. Unlike the standard
or like-particle RPA,
pnRPA provides means to calculate excited states of neighboring
isobars.
The starting point, however, is similar to the regular RPA: a mean-field
solution of the parent nucleus, which in turn defines a particle-hole basis. In our
case we began with a Hartree-Fock state $|\mathrm{HF} \rangle$, allowing
unrestricted deformation. The RPA can be derived a number of ways, but
it essentially approximates the energy surface about the mean-field state
as a harmonic oscillator. This leads to an RPA ground state $| \mathrm{RPA} \rangle$,
which implicitly includes zero-point fluctuations about the
mean-field state $|\mathrm{HF} \rangle$, and excited states are one-phonon excitations
on the ground state:
\begin{equation}
|\lambda\rangle=\beta_\lambda^\dagger |\mathrm{RPA}\rangle.
\end{equation}
For RPA one usually assumes the creation operators are simple
one-particle, one-hole operators. In order to properly discuss the
pnRPA, it is useful to further, if briefly, recapitulate the like-particle RPA
\cite{ring}. In the matrix formalism, one solves
\begin{equation}
\left (
\begin{array}{cc}
\mathbf{A} & \mathbf{B} \\
-\mathbf{B} & -\mathbf{A}
\end{array}
\right)
\left ( \begin{array}{c} \vec{X} \\ \vec{Y}  \end{array} \right) =
\Omega \left ( \begin{array}{c} \vec{X} \\ \vec{Y}  \end{array}
\right) \label{RPAEq}
\end{equation}
Here the matrix $\mathbf{A}$, a Hermitian matrix, is in fact the submatrix
of $H$ between one-particle, one-hole states (also
the matrix for the Tamm-Dancoff approximation), while the elements
of $\mathbf{B}$, a symmetric matrix, are the matrix elements of
$H$ between the HF state and two-particle, two-hole states.  (For
simplicity we assume real HF wavefunctions so that all subsequent
quantities are real.) One
can show that the excitation frequencies $\Omega$ are real
if the stability matrix,
\begin{equation}
\left (
\begin{array}{cc}
\mathbf{A} & \mathbf{B} \\
\mathbf{B} & \mathbf{A}
\end{array}
\right )
\end{equation}
has no negative eigenvalues; this is equivalent to the
Hartree-Fock state being at a (local) minimum. Part of the proof
of this requires that both $\mathbf{A} \pm \mathbf{B}$ have no
negative eigenvalues \cite{ring}, which in turn requires that
$\mathbf{A}$ have no negative eigenvalues. We emphasize this point
because the situation will be different for pnRPA.

One can show that the solutions of Eq.~(\ref{RPAEq}) come in
pairs: if $(\vec{X},\vec{Y})$ is a solution with frequency
$\Omega$, then $(\vec{Y},\vec{X})$ is also a solution with
frequency $-\Omega$. The solution with $\Omega > 0$ is associated
with the vector $(\vec{X},\vec{Y})$ such that $|X| > |Y|$, and one
chooses a normalization $|X|^2 - |Y|^2 = 1$; that one can do this
also derives from the nonnegative eigenvalues of the stability
matrix. The special case $\Omega = 0$ corresponds to invariance of
the Hamiltonian under a symmetry, for example, rotation or
translation; here the vector $(\vec{X},\vec{Y})$ cannot be
normalized, as $|X| =|Y|$, and one must resort to a different
formalism \cite{ring,stetcu2003,weneser}.

\begin{widetext}
\onecolumngrid

 The proton-neutron RPA is similar to the like-particle RPA but
with important differences. For like-particle RPA, the phonon creation
operator $\beta^\dagger$ uses one-particle, one-hole operators
of the form $\pi^\dagger \pi$, $\nu^\dagger \nu$ (thus the
name like-particle). The pnRPA phonon creation operator is also one-body
but changes the third component of isospin $T_z$:
\begin{equation}
\beta^\dagger_\lambda =
\sum_{mi} \left(X_{mi,\lambda}(pn)\pi^\dagger_m \nu_i -Y_{mi,\lambda}(np)
\nu^\dagger_i \pi_m\right) +\sum_{mi} \left(X_{mi,\lambda}(np)\nu^\dagger_m \pi_i -Y_{mi,\lambda}(pn)
\pi^\dagger_i \nu_m \right).
\label{cRPApn}
\end{equation}
The first and the fourth terms in this equation describe the
excited states of the $(Z+1,N-1)$ isobar, while the second and
third the $(Z-1,N+1)$ isobar. We follow the standard convention
and use indices $m,n$ for `particle' states, that is, unfilled
single-particle states in the Hartree-Fock wavefunction, and $i,j$
for `hole' states, or filled single-particle states; thus
$\pi^\dagger_m \nu_i$ destroys a filled neutron state (or creates
a neutron hole in the HF wavefunction, in the usual view) and
creates a proton in an excited particle state.
The equation of motion
\begin{equation}
\langle \mathrm{RPA} |[[\delta \beta,[H,\beta^\dagger]]| \mathrm{RPA}
\rangle = \Omega \langle \mathrm{RPA} |[\delta\beta,\beta^\dagger]|\mathrm{RPA}
\rangle,
\end{equation}
transforms as usual to a non-hermitian eigenvalue problem \cite{ring}
\begin{equation}
\left(\begin{array}{cccc}
A^{np,pn} & 0 & 0 & B^{np,pn} \\
0 & A^{pn,np} & B^{pn,np} & 0 \\
0 & -B^{np,pn} & -A^{np,pn} & 0 \\
-B^{pn,np} & 0 & 0 & -A^{pn,np}
\end{array}\right)\left(\begin{array}{c}
X(pn)\\
X(np)\\
Y(np)\\
Y(pn)\end{array}\right)=\Omega\left(\begin{array}{c}
X(pn)\\
X(np)\\
Y(np)\\
Y(pn)\end{array}\right),
\label{pnRPAeq}
\end{equation}
where the definitions for $A^{pn,np}$ and $B^{np,pn}$
matrices are similar to the regular proton-neutron conserving formalism,
where one approximates $ | \mathrm{RPA} \rangle \approx
| \mathrm{HF} \rangle$:
\begin{equation}
A_{mi,nj}^{np,pn}=\langle {\rm HF}|[\nu_i^\dagger \pi_m, [H,\pi_n^\dagger\nu_j]]|
{\rm HF} \rangle=(\epsilon^p_n-\epsilon_i^n)\delta_{mn}\delta_{ij}
-V^{pn}_{mn,ji},
\label{defA}
\end{equation}
\begin{equation}
B_{mi,nj}^{np,pn}=-\langle {\rm HF}|[\nu_m^\dagger \pi_i,[\pi_n^\dagger\nu_j,H]]|
{\rm HF} \rangle=-V_{in,jm}^{pn}.
\label{defB}
\end{equation}
(For the quasiparticle RPA (QRPA) one uses instead of the Hartee-Fock state
a Hartree-Fock-Bogoliubov state or a Bardeen-Cooper-Schrieffer state, and
the QRPA phonon is composed of quasiparticle-quasihole operators instead.)
The matrices $A^{pn,np}$ and $B^{pn,np}$ are defined similarly,
but are distinct unless $Z=N$; in fact, they have different dimensions
unless $Z=N$. Let $N_p^\pi$,
$N_h^\pi$ be number of proton particle and hole states, respectively, and
$N_p^\nu$, $N_h^\nu$ the number of neutron particle and hole states. Thus
the vectors $X(pn)$ and $Y(np)$ are of length $N_p^\pi N_h^\nu$ while
vectors $X(np)$, $Y(pn)$ are of length $N_p^\nu N_h^\pi$; the two lengths are
unequal unless $Z=N$. Similarly,
$\mathbf{A}^{np,pn}$ is a square matrix of dimension $N_p^\pi N_h^\nu$  while
$\mathbf{A}^{pn,np}$  is a square matrix of dimension $N_p^\nu N_h^\pi$,
while $\mathbf{B}^{np,pn}$ is a rectangular matrix of dimension $N_p^\pi N_h^\nu \times N_p^\nu N_h^\pi$,
and   $\mathbf{B}^{pn,np}= \left(\mathbf{B}^{np,pn}\right)^T$ .

\end{widetext}
\twocolumngrid

The zeroes in Eq.~(\ref{pnRPAeq}) occur because the original
Hamiltonian conserves charge (and thus $T_z$).
The overall form of Eq.~(\ref{pnRPAeq}) is identical
to that of Eq.~(\ref{RPAEq}); we have merely introduced a
block structure to $\mathbf{A}$, $\mathbf{B}$.  Because of
the zeroes in Eq.~(\ref{pnRPAeq}), the equations decouple:
\begin{equation}
\left(\begin{array}{cc}
A^{np,pn} & B^{np,pn} \\
-B^{pn,np} & -A^{pn,np} \\
\end{array}\right)\left(\begin{array}{c}
X(pn)\\
Y(pn)\end{array}\right)=\Omega\left(\begin{array}{c}
X(pn)\\
Y(pn)\end{array}\right)
\label{pnRPAd}
\end{equation}
and
\begin{equation}
\left(\begin{array}{cc}
A^{pn,np} & B^{pn,np} \\
-B^{np,pn} & -A^{np,pn} \\
\end{array}\right)\left(\begin{array}{c}
X(np)\\
Y(np)\end{array}\right)=\Omega\left(\begin{array}{c}
X(np)\\
Y(np)\end{array}\right)
\label{pnRPAe}
\end{equation}

It is important to note two things here. First, the decoupled
equations (\ref{pnRPAd}) and (\ref{pnRPAe}) are \textit{not} of
the same form as Eq.~(\ref{RPAEq}), because $\mathbf{A}^{pn,np}
\neq \mathbf{A}^{np,pn}$ unless $Z=N$. Because of this, some of
the usual theorems do not immediately apply to (\ref{pnRPAd}) and
(\ref{pnRPAe}), especially regarding the positivity of $\Omega$
(more about this, however, in a moment) and the sign of
$|X|^2-|Y|^2$.

Second, any solution $(X(pn),Y(pn))$ to Eq.~(\ref{pnRPAd}) with
frequency $\Omega$ is related to a solution $(X(np),Y(np))$ of
Eq.~(\ref{pnRPAe}) with frequency $-\Omega$, by $(X(np),Y(np)) =
(Y(pn),X(pn))$, up to some overall phase factor. This simplifies
finding solutions.  Let $(U,V)$ be a solution of
Eq.~(\ref{pnRPAd}) with frequency $\omega$, which can be positive
or negative.  If $|U|^2 -|V|^2 > 0$, and so normalizable to $1$,
then let $(X(pn),Y(pn)) =(U,V)$ with frequency $\Omega = \omega$.
Otherwise, if $|U|^2 - |V|^2 < 0$, then let $(X(np),Y(np))=(V,U)$
with frequency $\Omega = -\omega$ be the solution to
Eq.~(\ref{pnRPAe}). In both cases one can normalize to 1.

What about $\Omega = 0$?
In like-particle RPA, if exact symmetries are broken by the mean-field,
one obtains zero eigenvalues; the corresponding eigenvectors,
which identify the generators of the broken symmetries \cite{thouless61},
are not normalizable and must be treated with care. For pnRPA, although zero
eigenvalues are not excluded, the corresponding eigenvectors do not
play a special role and are in fact normalizable, as argued below.

In like-particle RPA, one can show that stability of the
mean-field state implies that $\Omega \geq 0$.  But now in pnRPA
one can have $\Omega < 0$, even for $|X|^2 - |Y|^2 =1$. At first
glance this is troubling; however, we provide two arguments which
resolve this apparent paradox.

First, because pnRPA allows charge-changing phonons of the form
(\ref{cRPApn}), one should, at least implicitly, also perform the
Hartree-Fock minimization allowing mixing of proton and neutron
states, with $N$-$Z$ fixed by a Lagrange multiplier $\lambda$. (We
have in fact done such a calculation, but the results are
indistinguishable from fixing $N$-$Z$ by hand, that is, as one
varies $\lambda$, $N$-$Z$ makes integral jumps. Nonetheless, the
image is useful.)  Now the constrained Hartree-Fock is at a true
minimum, and if one considers the eigenfrequencies of
(\ref{pnRPAeq}) all the resultant pnRPA frequencies are proven
positive-definite (although in practice one solves (\ref{pnRPAd})
or (\ref{pnRPAe}) instead). To interpret the results as physical
transition energies, however, one has to subtract off
$\lambda(N-Z)$ which leads to negative frequencies.

One can see this directly. By adding the Langrange multiplier,
one shifts the  diagonals of the $\mathbf{A}$
matrices,
\begin{equation}
\tilde A^{np,pn}= A^{np,pn}-\lambda (N-Z) \cdot I,
\label{Ashift1}
\end{equation}
and
\begin{equation}
\tilde A^{pn,np}= A^{pn,np}+\lambda (N-Z) \cdot I.
\label{Ashift2}
\end{equation}
By substituting  into Eq.~(\ref{pnRPAd}), it is clear
one is adding a constant along the entire diagonal, which
only shifts  $\Omega$ to $\tilde \Omega = \Omega + \lambda(N-Z)$,
while the frequencies in Eq.~(\ref{pnRPAe}) are shifted in the opposite
direction.  In fact, for \textit{any} value of $\lambda$
one only shifts the frequencies and the solutions $(X,Y)$ are manifestly
unchanged.

We call this result the shift lemma: by adding the Lagrange
multiplier one shifts the relative position of the proton and
neutron Fermi surfaces, but the only result is shifting the
absolute value of the pnRPA frequencies. The relative frequencies
and the eigenvectors are unchanged. There is a useful consequence:
if one obtains a zero mode, from the shift lemma one can find a
case where the frequency corresponding to the same eigenvector is
positive, and one expects the pnRPA vector to be normalizable; and
as the eigenvector is independent of the shift, it is
\textit{always} always normalizable. The reader should note that
the shift lemma only arises because of the unique isospin
dependence of the block decomposition (\ref{pnRPAeq}); no such
shift is possible for like-particle RPA because one cannot add a
constant to the diagonal.

\section{Results}

We test the reliability of pnRPA's predictions for GT strengths
against exact diagonalization in full $0\hbar\omega$ SM spaces.
That is, we calculate the GT strength distributions for several
nuclei in the $sd$ ($1s_{1/2}$-$0d_{3/2}$-$0d_{5/2}$) shell on top
of an inert $^{16}$O core, and two Ti isotopes in the $pf$
($1p_{1/2}$-$1p_{3/2}$-$0f_{5/2}$-$0f_{7/2}$) shell above a
$^{40}$Ca core. While we do not compare directly with experiment,
we use in our calculations phenomenological interactions which are
very successful in reproducing the experimental data: Wildenthal
`USD' in the $sd$ shell \cite{wildenthal}, and Richter-Brown in
the $pf$ shell \cite{RvJB91}. The shell model diagonalization were
performed using a descendent of the Glasgow code
\cite{Whitehead77}, and the shell-model strength distributions
using an efficient Lanczos moment method \cite{LanczosMoment}.

\begin{widetext}
\onecolumngrid

Having defined the pn-phonon creation operator $\beta^\dagger $ in (\ref{cRPApn}), one can
calculate the transition matrix element required by Eq. (\ref{trStr}) as
\begin{equation}
\langle \mathrm{RPA} | {\cal O}|\lambda_{(Z\pm 1,N\mp 1)}\rangle=\langle \mathrm{RPA} |
[{\cal O},\beta_\lambda^\dagger]|\mathrm{RPA}\rangle = \sum_{mi}\left(X_{mi}^\lambda(pn/np)
{\cal O}_{mi}+Y_{mi}^\lambda(pn/np){\cal O}_{im}\right).
\end{equation}
In our case, ${\cal O}$ is the GT transition operator defined in (\ref{defOp}),
which induces transitions between the correlated ground state of the parent nucleus and
the $(Z+1,N-1)$ or $(Z-1,N+1)$ isobars.
\end{widetext}
\twocolumngrid

Table \ref{GT} summarizes results  for total transition strengths, centroids
and distribution widths, where indeed we find that the pnRPA
moments are reasonably close to SM. As a check, note that
the Ikeda sum rule (\ref{IkedaSR}) is fulfilled in both SM and pnRPA, as expected.
We find same features typical of like-particle RPA,
that is the pnRPA centroids are usually lower in energy, while the SM distribution
widths are larger \cite{Goeke82,stetcu2003}.
The latter can be understood as particle-hole correlation
beyond RPA which further fragment the distributions.

Note that we have results not only for even-Z, even-N nuclides but
also odd-odd and odd-A, all with comparable success.
(A technical note: In our mean-field calculations, we allow only real wave functions.
This does not have any effect for the even-even nuclei. For even-odd or odd-$A$ nuclei however,
the exact mean-field solution could be complex, and we see small symptoms
of this restriction:
because the rotations with respect to $x$ and $z$ axes are complex, the proton
and neutron number conserving RPA formalism identifies some generators of the broken
symmetries as lying at small, but \textit{not} zero, excitation energies.
While these approximations can be relevant for other transitions \cite{stetcu2003},
they do not have an impact here.)

\begin{table}
\caption{Total strength $S_0$, centroid $\bar E$, and width
$\Delta E$ for GT transition operator. The nuclei have been
grouped into even-even, odd-odd and odd-$A$.} \label{GT}
\begin{ruledtabular}
\begin{tabular}{cccccccc}
 & & \multicolumn{2}{c}{$S_0$} & \multicolumn{2}{c}{$\bar E$ (MeV)}
 & \multicolumn{2}{c}{$\Delta E$ (MeV)}\\
\cline{3-4}\cline{5-6}\cline{7-8}
Nucleus & & SM & RPA & SM & RPA & SM & RPA  \\
\hline
$^{20}$Ne & $\beta^+/\beta^-$ & 0.55 & 0.69 & 15.81 & 12.20 & 4.22 & 2.42 \\
$^{22}$Ne & $\beta^+$ & 0.50 & 0.63 & 19.71 & 16.17 & 3.81 & 1.33 \\
          & $\beta^-$ & 6.50 & 6.63 & 4.48  &  4.75 & 5.64 & 3.79 \\
$^{24}$Ne & $\beta^+$ & 0.51 & 0.61 & 19.73 & 18.37 & 3.31 & 1.36 \\
          & $\beta^-$ &12.51 &12.61 &  3.82 &  4.26 & 4.91 & 3.46 \\
$^{24}$Mg & $\beta^+/\beta^-$ & 2.33 & 2.73 & 13.40 & 10.92& 3.86 & 2.33 \\
$^{26}$Mg & $\beta^+$ & 1.78 & 2.05 & 15.72 & 13.42 & 3.55 & 1.97 \\
          & $\beta^-$ & 7.78 & 8.05 &  6.94 & 5.93 & 5.58 & 4.62 \\
$^{28}$Si & $\beta^+/\beta^-$ & 3.89 & 3.39 & 13.54 & 12.29& 3.07 & 1.86 \\
$^{30}$Si & $\beta^+$ & 2.52 & 2.33 & 15.59 & 13.39 & 2.59 & 1.83 \\
          & $\beta^-$ & 8.52 & 8.33 &  8.69 &  7.38 & 4.69 & 3.26 \\
$^{32}$S  & $\beta^+/\beta^-$ & 4.01 & 4.25 & 12.48 & 10.38& 3.04 & 2.21\\
$^{34}$S  & $\beta^+$ & 1.59 & 1.88 & 14.22 & 11.90 & 2.53 & 2.13 \\
          & $\beta^-$ & 7.59 & 7.88 & 7.91  &  7.88 & 4.06 & 2.34 \\
$^{36}$Ar & $\beta^+/\beta^-$ & 2.10 & 2.22 & 12.09 & 10.07& 2.59 & 2.57\\
$^{44}$Ti & $\beta^+/\beta^-$ & 0.61 & 0.79 & 9.95  &  7.96& 2.27 & 1.50\\
$^{46}$Ti & $\beta^+$ & 0.44 & 0.60 & 12.47 & 10.46 & 1.80 & 0.55 \\
          & $\beta^-$ & 6.44 & 6.60 & 2.99  & 3.31  & 3.51 & 2.39 \\
\hline
$^{24}$Na & $\beta^+$ & 1.67 & 1.92 & 14.59 & 12.34 & 3.53 & 2.65 \\
          & $\beta^-$ & 7.67 & 7.92 & 6.67 & 6.15   & 4.87 & 3.72 \\
$^{26}$Al & $\beta^+/\beta^-$ & 4.28 & 4.28 & 11.86 & 10.37 & 3.43 & 2.87\\
\hline
$^{21}$Ne & $\beta^+$ & 0.63 & 0.67 & 15.85 & 13.96 & 4.49 & 2.82 \\
          & $\beta^-$ & 3.63 & 3.67 & 6.49  &  5.82 & 5.05 & 4.17 \\
$^{25}$Na & $\beta^+$ & 1.39 & 1.50 & 15.96 & 14.06 & 3.27 & 1.95 \\
          & $\beta^-$ &10.39 & 10.50& 5.27  &  4.92 & 5.13 & 3.97 \\
$^{27}$Al & $\beta^+$ & 3.20 & 2.52 & 14.04 & 12.72 & 2.94 & 2.02 \\
          & $\beta^-$ & 6.20 & 5.52 &  9.33 &  8.61 & 5.20 & 3.75 \\
$^{29}$Al & $\beta^+$ & 1.80 & 1.97 & 16.10 & 13.51 & 2.76 & 2.04 \\
          & $\beta^-$ &10.80 &10.97 &  6.61 &  6.45 & 4.85 & 2.99
\end{tabular}
\end{ruledtabular}
\end{table}

Figs.~\ref{ne24}--\ref{s32}  illustrate selected results in detail.
These are typical results, neither better nor worse on the average.
These figures follow a useful convention used by many many authors and
plot the accumulated sum of the strength, $\sum B(GT) = \sum_{E < E_f} S(E)$ , which allows
one to compare by eye the first few moments.
Again we see by eye generally good results, for even-even, odd-odd, and
odd-A alike.  The major systematic error of RPA appears to be lower centroids
for $\beta^+$ transitions; why there is no similar lowering of the $\beta^-$
centroid is unknown.

In Figs.~\ref{ne24} and \ref{mg26} we compare $\beta^+$
calculation against the spherical pnQRPA calculations of Lauritzen
\cite{Lau88}, which are similar to those of \cite{ZB93,ABB93}.
(All of these papers computed only $\beta^+$ decay, and only for
even-even nuclides.) In general the QRPA strengths $S_0$ are about
twice as much as the exact calculation, and the centroids
$\bar{E}$ are significantly lower. The calculations of
\cite{ABB93} found that pnQRPA gives results very similar to
shell-model calculations restricted to 2 particle, 2 hole
excitation out of the $0d_{5/2}$ orbit. This is believeable, as
RPA correlations are approximately 2 particle, 2 hole in nature.
By contrast, our pnRPA calculations on top of a deformed HF state
do much better. In particular, our deformed pnRPA calculations
better approximate the correct total strength than does the
spherical pnQRPA.
 That deformation
can ``quench'' Gamow-Teller strength is already known from shell
model calculations \cite{AZZ93,TDH96}. (See, however, pnQRPA
calculations of \cite{Sar03}, who find the total strength to be
insensitive to deformation, although the \textit{details} of the
distribution they find to be sensitive. They speculate that the
difference is that they include up to 10 harmonic oscillator
shells in their calculations, while our calculations and those of
\cite{AZZ93} and \cite{TDH96} are within a single harmonic
oscillator shell. Why a multi-shell calculation should be
insensitive to deformation is not clear.)

There is a curious exception: the spherical pnQRPA calculation of
\cite{CMS91} on $^{26}$Mg yields $S_0$ and $\bar{E}$ with
approximately the same accuracy relative to SM results as us. They
only compute the one $\beta^+$ decay, however, and it is difficult
to understand the difference between their methodology and that of
\cite{Lau88,ZB93,ABB93}; furthermore, although in principle all
three calculations are using the same shell-model interaction
\cite{wildenthal}, which we properly scaled to $A=26$, we can
reproduce Lauritzen's shell model results \cite{Lau88} but not
those of \cite{CMS91}.

In order to further illustrate how important it is to treat
correctly deformation, we show in Fig. \ref{s32} three different
pnRPA calculations for $^{32}$S: with dashed line we represent the
distribution strength for a spherical HF state, with dotted line,
a distribution obtained by starting with an oblate HF state and
with dash-dotted line a distribution obtained starting with a
triaxial HF state. The two deformed states are almost degenerate
in energy, the triaxial state being slightly lower. From
Fig.~\ref{s32} one sees that the strength distribution off the
triaxial HF state best approximates the SM result (as one might
expect). QRPA strength distributions which use a starting
spherical mean field solution \cite{Lau88} are slightly more
fragmented than in our spherical calculation (we obtain just one
state with $J=1$), but the total strength is about the same in
both pnRPA and QRPA, overestimating seriously the SM total
strength. We conclude that using a deformed mean-field solution is
at least as important, and arguably more so, than treating pairing
with rigor.

In the above discussion we have focussed on the gross properties of the
Gamow-Teller ``resonance,'' which is the main application of RPA.
But for cold systems, one is often most interested exclusively in low-lying
transitions. As is observable in the figures, our pnRPA calculations
are rather mediocre when it comes to the lowest-lying GT transitions:
arguably better than pnQRPA, but not very good compared to the exact
shell model results.  This is an important caveat for any applications.

\begin{widetext}
\onecolumngrid

\begin{figure}
\centering
\includegraphics*[scale=0.6]{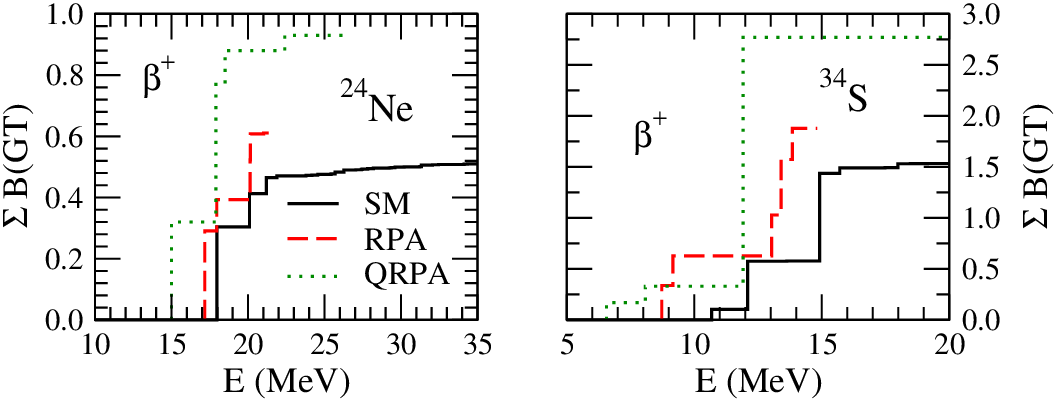}
\caption{SM (full curve), RPA (dashed),
and QRPA(dotted) summed strength
for $\beta^+$ (left) GT transitions from
$^{24}$Ne, $^{34}$S. The
QRPA calculation is from \cite{Lau88} }
\label{ne24}
\end{figure}

\begin{figure}
\centering
\includegraphics*[scale=0.6]{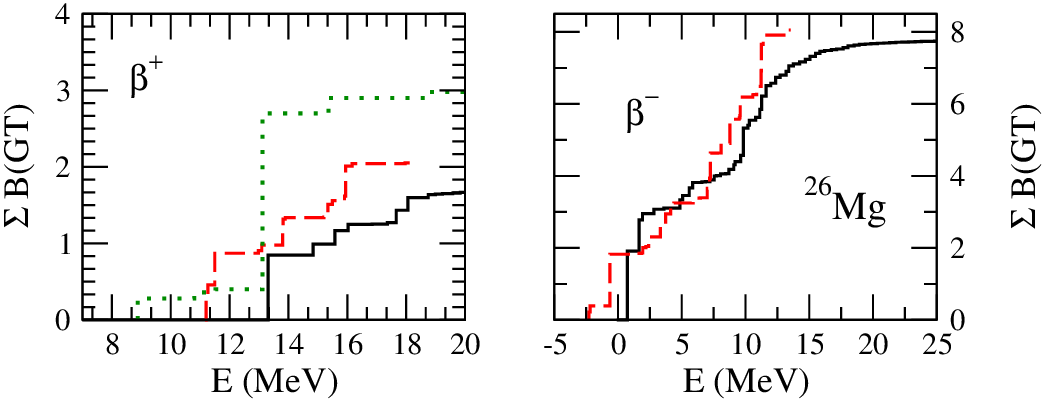}
\caption{SM (full curve) and RPA (dashed curve) summed strength
for $\beta^+$ (left) and $\beta^-$ (right) GT transitions in
$^{26}$Mg. For $\beta^+$ transitions we also include the
QRPA calculation \cite{Lau88} (dotted curve)}
\label{mg26}
\end{figure}

\begin{figure}
\centering
\includegraphics*[scale=0.6,angle=-90]{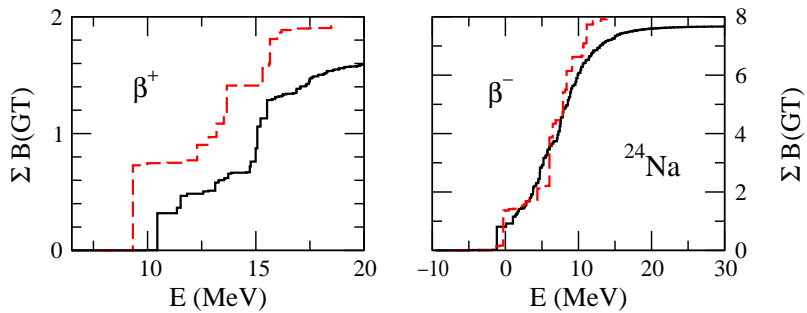}
\caption{SM (full curve) and RPA (dashed curve) summed strength
for $\beta^+$ (left) and $\beta^-$ (right) GT transitions in
$^{24}$Na.}
\label{na24}
\end{figure}

\begin{figure}
\centering
\includegraphics*[scale=0.6,angle=-90]{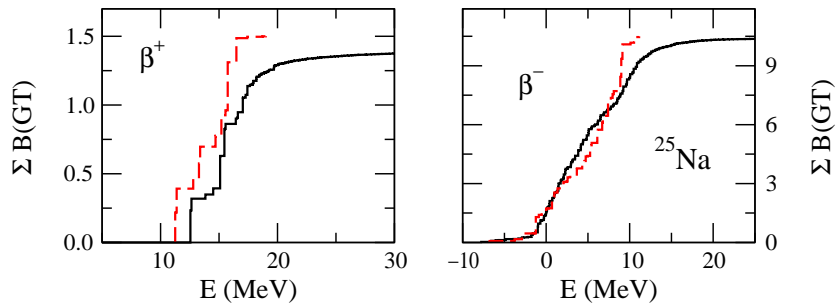}
\caption{Same as in Fig. \ref{na24}, but for $^{25}$Na.}
\label{na25}
\end{figure}

\begin{figure}
\centering
\includegraphics*[scale=0.6]{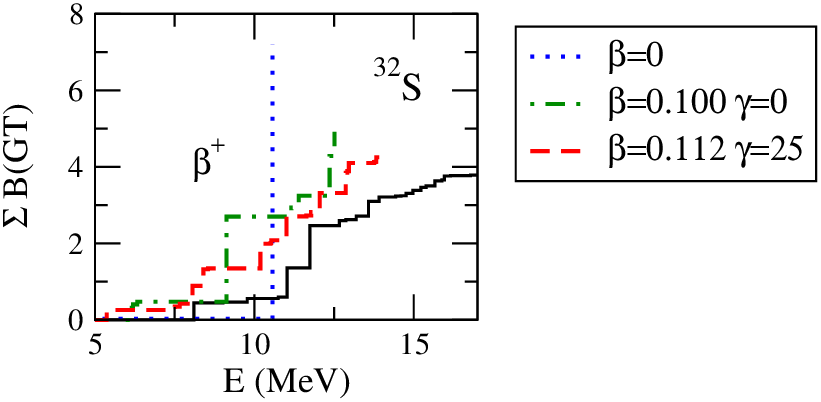}
\caption{Comparison of the SM (full curve) and pnRPA
predictions for the summed strength of GT $\beta^+$ transitions in $^{32}$S.
The pnRPA was performed on top of a spherical
(dotted curve), oblate (dash-dotted), and triaxial (dashed)
HF state respectively, emphasizing the importance of a correct treatment of
deformation for a correct description of the strength distribution.}
\label{s32}
\end{figure}

\end{widetext}
\twocolumngrid

\section{Conclusions}

The main purpose of this paper was to investigate the pnRPA's reliability
for predicting $\beta^\pm$ GT transition strengths; the motivation is
astrophysical applications, where, aside from binding energies and
electomagnetic
transitions, a good knowledge of week transitions is essential.
Our tool in this investigation was the interacting SM which can provide
the exact numerical solution in a restricted space.
Although we made our tests for nuclides near the bottom of the
valley of stability, presumably our results apply out to the driplines;
in fact the major uncertainty will be the shell-model interaction,
not the pnRPA.

We show a very good agreement between SM and pnRPA $\beta^\pm$ transition
distributions
in a large number of nuclides in the $sd$ and $pf$ shells, similar to
high-lying
collective electromagnetic transitions investigated
in a previous paper \cite{stetcu2003}. Furthermore, we obtain
 better
results than spherical pnQRPA, obtaining better suppression of the
total strength. This may seem  surprising, as we do not treat
rigorously the pairing interaction; on the other hand, we model
correctly the deformation and this proves suitable for describing
GT transition strengths.
(Because of this, we view tests of pnRPA in schematic models
without symmetry breaking \cite{SMS01} as being inadequate.)
 It is possible that a deformed pnQRPA
approach, with better treatment of pairing, will lead to further
improvement.  We hope however to extend our program to the
deformed pnQRPA in the not-so-distant future.  Furthermore, it will
be important to consider multi-shell spaces, where at least one set
of deformed pnQRPA calculations (not validated by direct shell model
calculations, however, due to the enormous size of the model space) suggest
a weaker dependence on deformation \cite{Sar03}.

\begin{acknowledgments}
We would like to thank J.~Engel for helpful conversations
regarding the stability matrix and the reality of the pnRPA
frequencies, and J.~Hirsch for pointing out some useful references.
The U.S.~Department of Energy supported this
investigation through grant DE-FG02-96ER40985, and the
U.S.~National Science Foundation through grant PHY-0140300.

\end{acknowledgments}

\end{document}